\documentclass[preprint,12pt,3p]{elsarticle}
\makeatletter
\def\ps@pprintTitle{%
 \let\@oddhead\@empty
 \let\@evenhead\@empty
 \def\@oddfoot{}%
 \let\@evenfoot\@oddfoot}
\makeatother

\usepackage{amsmath}
\usepackage{amssymb}
\usepackage[]{hyperref}

\begin{document}

\begin{frontmatter}

\title{Self-similarity and vanishing diffusion in fluvial landscapes} 

\address[label1]{Department of Civil and Environmental Engineering, Princeton University, Princeton, NJ 08544}
\address[label2]{High Meadows Environmental Institute, Princeton University, Princeton, NJ 08544}
\address[label3]{Department of Mathematics, Stony Brook University, Stony Brook, NY 11794}

\cortext[cor1]{I am corresponding author}

\author[label1]{Shashank Kumar Anand\corref{cor1}}
\ead{skanand@princeton.edu}
\author[label2]{Matteo B. Bertagni}
\author[label3]{Theodore D. Drivas}
\author[label1,label2]{Amilcare Porporato\corref{cor1}}
\ead{aporpora@princeton.edu}

\begin{abstract}
Complex topographies exhibit universal properties when fluvial erosion dominates landscape evolution over other geomorphological processes. Similarly, we show that the solutions of a minimalist landscape evolution model display invariant behavior as the impact of soil diffusion diminishes compared to fluvial erosion at the landscape scale, yielding complete self-similarity with respect to a dimensionless channelization index. Approaching its zero limit, soil diffusion becomes confined to a region of vanishing area and large concavity or convexity, corresponding to the locus of the ridge and valley network. We demonstrate these results using 1D analytical solutions and 2D numerical simulations, supported by real-world topographic observations. Our findings on the landscape self-similarity and the localized diffusion resemble the self-similarity of turbulent flows and the role of viscous dissipation. Topographic singularities in the vanishing diffusion limit are suggestive of shock waves and singularities observed in nonlinear complex systems.
\end{abstract}


\end{frontmatter}


\section{Introduction}
\label{S1}
The dominance of fluvial erosion leads to the emergence of self-similarity in natural landscapes, as the interlocked network of ridges and valleys grows in complexity. Statistical self-similarity in such landscapes reveals invariant statistical structures across different observation scales \cite{tarboton1988fractal, rinaldo1993self, rigon1996hack, rodriguez2001fractal, banavar2001scaling, kovchegov2022random}, giving rise to scaling laws for several key properties like contributing area, stream length, and drainage density, as the drainage networks become fractal \cite{huang1989fractal, rodriguez2001fractal, rinaldo2014evolution}. Such universal scaling laws suggest that landscape dynamics eventually become independent of the precise fluvial erosion intensity, attaining complete self-similarity with respect to the fluvial erosion process \cite{barenblatt1996scaling, porporato2022hydrology}. 

Several modeling studies have used dimensional analysis to identify the fundamental groups that control the emergence of similarity solutions of landscape evolution models (LEMs) \cite{willgoose1991coupled, whipple1999dynamics, simpson2003topographic, Perron2009, bonetti2020channelization}. In cases where a single length scale characterizes the domain geometry, the landscape dynamics is controlled by a dimensionless channelization index ($\mathcal{C_I}$) \cite{anandcomment2022}. The channelization index quantifies the relative balance between fluvial erosion and soil diffusion in transporting the sediment influx through uplift, thereby describing the tendency to form complex ridge/valley networks. 

The first goal of this study is to explore the self-similarity in the solutions of a simple landscape evolution model at large channelization indices ($\ln{\mathcal{C_I}} \gg 1$). At such $\mathcal{C_I}$ values, the influence of soil diffusion on landscape evolution diminishes compared to fluvial erosion globally (i.e., across the entire landscape domain). Previous investigations have shown that local landscape properties, like the mean elevation profile and power spectra, show self-similarity at large $\mathcal{C_I}$ \cite{hooshyar2020pre, hooshyar2021grl}. Recently, it has been suggested that global landscape properties could also reach complete self-similarity for large channelization index values \cite{porporato2022hydrology}. The emergence of self-similarity is not uncommon in complex systems \cite{carr1999self, eggers2008role, goldenfeld2018lectures} and here presents intriguing parallels with fully developed turbulence \cite{barenblatt2014turbulent, frisch1995turbulence, smits2011high}.

The second question pertains to the role of diffusion (i.e., soil creep \cite{culling1963soil}) in fluvial landscapes. Although the impact of diffusion diminishes globally over a landscape for asymptotically large $\mathcal{C_I}$ values, we show that, in reality, it persists localized to regions with sharp curvatures. Batchelor \cite{batchelor1956steady} articulated a similar localized stabilization by viscous dissipation in high Reynolds number flows: `viscous forces act on the fluid as small everywhere, except perhaps in the neighborhood of certain surfaces in the fluid.' Since then, the question of the role of vanishing diffusion in the Navier-Stokes equations (or equivalently, the existence of viscosity solutions in the fluid dynamic Euler equations) has remained essential \cite{kato1972nonstationary, lopes2008vanishing, drivas2019remarks, eyink2022onsager}. 

From a formal point of view, the vanishing diffusion solutions of LEMs correspond to viscosity solutions of the stream power equation (i.e., LEM formally with no soil diffusion \cite{whipple1999dynamics, willett1999orogeny, Perron2009}) in the sense of the viscosity approach introduced by Crandall and Lions \cite{crandall1983viscosity, rouy1992viscosity} to get unique weak solutions of Hamilton-Jacobi equations for the control problem. The sharp fronts of the landscape slope are thus analogous to discontinuity formation in surface growth models \cite{kardar1986dynamic, qi2001self} and generation of shock waves \cite{bec2007burgers, whitham2011linear, kluwick2018shock}. 

\section{Governing equations and role of channelization index}

We consider a minimalist landscape evolution model (LEM) for the evolution of surface elevation $z$ and specific contributing area $a$ \cite{bonetti2020channelization}. The elevation dynamics is controlled by tectonic uplift, soil creep (i.e., diffusion), and fluvial erosion \cite{howard1994detachment, Perron2009, bonetti2020channelization}. Tectonic uplift ($U$) is the forcing acting beneath the surface and is usually modeled as a constant source term. Soil diffusion ($D \nabla^2 z$, $D$ is the diffusion coefficient) represents the effect of various biophysical processes to smooth the topography \cite{culling1963soil}. Detachment-limited fluvial erosion is proportional to the shear stress by the flowing runoff over the surface and is modeled as a sink term ($K a^m |\nabla z|^n$, where $K$ is the erosion coefficient, $m$ and $n$ are model exponents) such that the eroded sediments do not get re-deposited inside the domain \cite{whipple1999dynamics, anand2022inception}.

The governing equation for the specific contributing area is derived from the water continuity equation, describing quasi-steady flow down the surface slope resulting from unitary runoff-producing precipitation \cite{bonetti2018theory, anand2022inception}. We focus on a simple geometry, consisting of a long strip of width $l$ and fixed elevation ($z=0$) at the boundaries (except in the analogy with the Burgers vortex, where the length scale is given by $\sqrt{D/K}$). The initial condition for elevation consists of small random spatial noise, where any local minimum inside the domain is filled due to the singular character of $a$ \cite{bonetti2018theory}. The equation for $a$ is a free boundary problem, where the boundary is the set of critical points (i.e., local maxima and saddles of $z$ in the domain interior). The latter is not determined a priori but rather dynamically found while solving the equation for $z$.

Using $K$, $U$, and $l$ as scaling variables, the dimensionless governing equations are
\begin{eqnarray}
& \frac{\partial \hat{z}}{\partial \hat{t}} = \frac{1}{\mathcal{C_I}}\hat{\nabla}^2 \hat{z} - \hat{a}^m | \hat{\nabla} \hat{z}|^n + 1, \label{eq:z_nd} \\
& - \hat{\nabla} \cdot \left(\hat{a}\, \frac{\hat{\nabla} \hat{z}}{|\hat{\nabla} \hat{z}|} \right) = 1, \label{eq:a_nd}
\end{eqnarray}
where $\left(\hat{\cdot}\right)$ denotes the dimensionless form of the involved physical quantities. $\mathcal{C_I}$ is the channelization index
\begin{eqnarray}
\label{eq:CI}
\mathcal{C_I} = \frac{K^{\frac{1}{n}}l^{\frac{m}{n}+1}}{DU^{\frac{1}{n}-1}},
\end{eqnarray}
which presents the relative balance between fluvial erosion and diffusion in transporting the sediment influx by uplift and quantifies the landscape tendency to form channels. Similar dimensionless quantities have been derived in previous studies \cite{howard1994detachment, Perron2009, mcguire2016controls, bonetti2020channelization, anand2020ems}, which vary in their formulation based on the scaling variables used for the dimensional analysis. $\mathcal{C_I}$ provides a measure of dynamic similitude for given $m$ and $n$, namely two distinct model configurations are dynamically equivalent if they have the same $\mathcal{C_I}$ value \cite{anandcomment2022}. The role of the channelization index is reminiscent of the global Reynolds number $Re$ in fluid dynamics, which represents the ratio of advective to viscous forces and describes the flow tendency to become turbulent (\ref{app:CI} for details).

\section{Complete self-similarity in the fluvial erosion regime}

To assess the existence of a self-similar regime for the global landscape properties at large $\mathcal{C_I}$, we focus on the sediment flux transported out of the domain by fluvial erosion and soil diffusion. For a steady topography, the balance between fluvial erosion and diffusion sediment outflux can be calculated by integrating Eq. \eqref{eq:z_nd} over the domain (see \ref{app:gsb}) as
\begin{equation}
\label{glo_bud_nd_2}
\underbrace{\frac{2}{\mathcal{C_I}}
\hat{S_b}}_{\text{Outflux by Diffusion}} + \underbrace{\langle \hat{E}_{DL}  \rangle \vphantom{\frac{1 }{\mathcal{C_I}}}}_{\text{Outflux by Fluvial Erosion}}= \underbrace{\vphantom{ \frac{1}{\mathcal{C_I}}} 1,}_{\text{Influx by Uplift}}
\end{equation}
where $\hat{S_b}$ is the average boundary slope (\ref{app:simset}, Fig. \ref{fig:numerical_setup}) and $\langle \hat{E}_{DL} \rangle$ is the average fluvial outflux. 

Eq. \eqref{glo_bud_nd_2} shows that at a given $\mathcal{C_I}$ value, the amount of sediment transported out of the domain by soil diffusion in comparison to fluvial erosion is characterized by dimensionless $\hat{S_b}$. As a result, one can write a relationship for $\hat{S_b}$ as a function of the channelization index and the fluvial exponents, namely
\begin{equation}
\label{eq:phi_1}
    \hat{S_b} = \varphi_1 \left( \mathcal{C_I}, m, n \right),
\end{equation}
where the functional form is obtained by analyzing the solutions of Eqs. \eqref{eq:z_nd} and \eqref{eq:a_nd}.

\begin{figure*}[!ht]
\centering
\includegraphics[width=0.86\linewidth]{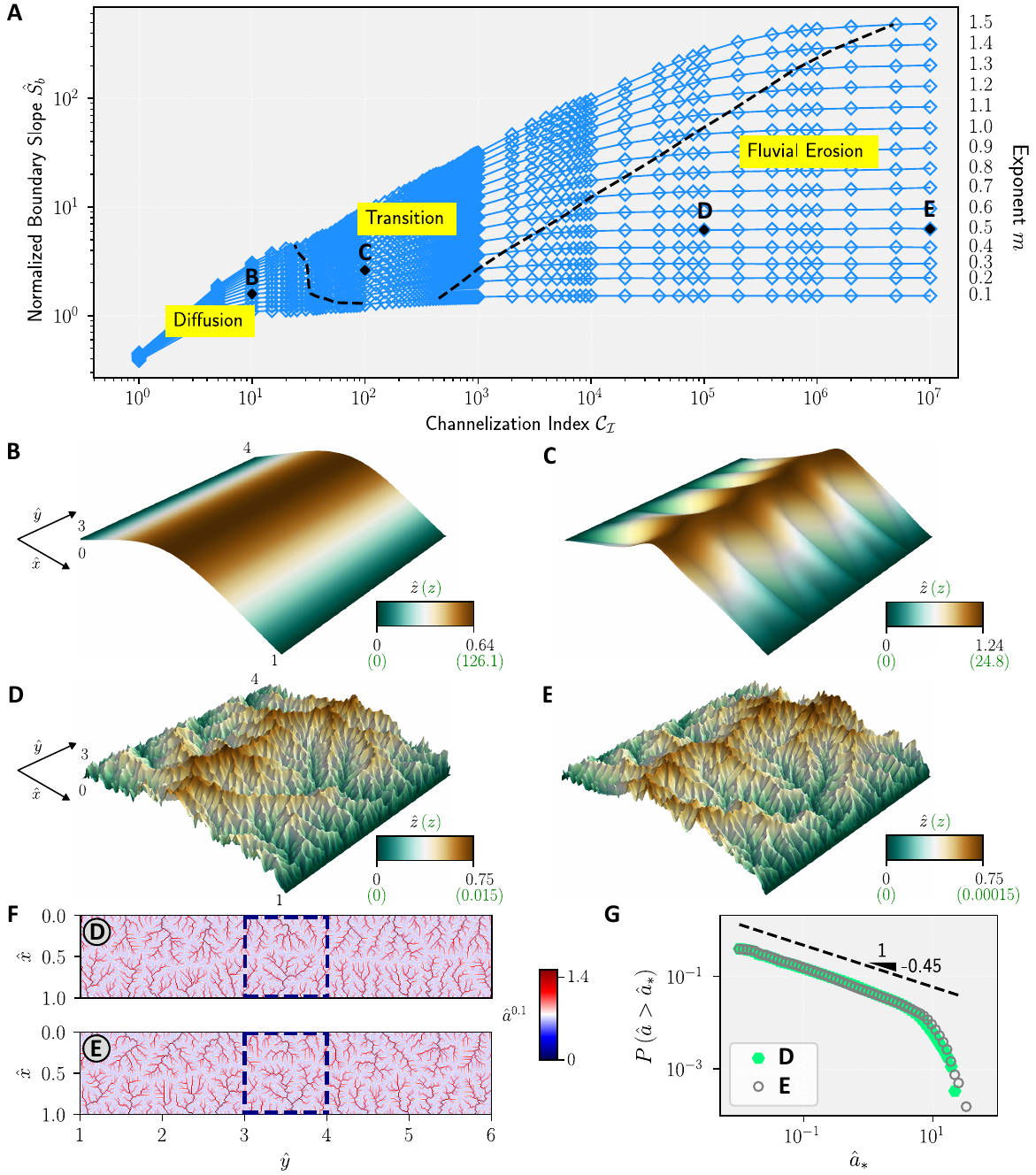}
\caption{\label{fig:one}Emergent self-similarity at asymptotically large channelization-index values. (A) $\hat{S}_b$ as a function of $\mathcal{C_I}$ for $m$ from 0.1 to 1.5, keeping $n=1$. The dashed curves separate the diffusion, transition, and fluvial erosion regimes. $\hat{S}_b$ reaches a self-similar plateau as a function of $m$ in the fluvial erosion regime. This landscape transition is reminiscent of the transition from laminar flow to fully developed turbulence in a pipe for asymptotically large Reynolds number values. (B-E) Steady-state landscapes for $m=0.5$; also marked in A. (B,C) Landscapes at $\mathcal{C_I} = 10$ (diffusion regime) and $\mathcal{C_I} = 10^2$ (transition regime), with different dimensional ($z$) and dimensionless ($\hat{z}$) elevation fields. Self-similar landscapes from fluvial erosion regime at (D) $\mathcal{C_I} = 10^5$ and (E) $\mathcal{C_I} = 10^7$ have different dimensional elevation fields ($z$), but similar dimensionless elevation fields ($\hat{z}$). (F) The specific contributing area $\hat{a}$ shows the flow accumulation in landscapes from D and E. The dashed rectangles denote the regions shown in D and E. (G) Exceedance probability distributions of $\hat{a}$ for the landscapes in F exhibit a remarkable collapse with a power-law scaling exponent $\approx -0.45$.} 
\end{figure*}

Detailed numerical simulations reveal that the global sediment budget (Fig. \ref{fig:one}A) is characterized by three distinct regimes as a function of the channelization index (please refer to \ref{app:vertest} for details on the numerical verification tests). In the diffusion regime at low $\mathcal{C_I}$ values, the topography remains smooth (e.g., Fig. \ref{fig:one}B) as any surface instability is smeared out by soil diffusion and $\hat{S}_b$ increases with $\mathcal{C_I}$. As the channelization index grows beyond a critical threshold (dashed curve in Fig. \ref{fig:one}A), fluvial erosion becomes strong enough to form valleys \cite{anand2022inception}, initiating the transition regime, where $\hat{S}_b$ remains a function of $\mathcal{C_I}$. A typical landscape from this regime is shown in Fig. \ref{fig:one}C. As $\mathcal{C_I}$ further increases and the ridge/valley network becomes more intricate, $\hat{S}_b$ reaches a plateau (invariant with respect to $\mathcal{C_I}$), defining the fluvial erosion regime. The curve separating the fluvial erosion regime from the transition regime can be approximated as an exponential function, $\mathrm{e}^{\left( m + 0.89 \right)/0.17}$, for $m \in \left(0.1,1\right)$.

In the classification of self-similar problems \cite{barenblatt1996scaling, porporato2022hydrology}, the self-similarity in the fluvial erosion regime
\begin{equation}
\label{eq:phi_2}
\hat{S_b} = \varphi_2\left(m,n\right)
\end{equation}
is said to be complete because the function in Eq. \eqref{eq:phi_2} reaches a finite value for $\ln{\mathcal{C_I}} \gg 1$. Numerical results for a range of $m$ and $n$ values are given in the \ref{app:diffmn} and \ref{app:varp}, where the connection of these results to the optimality principle for fluvial landscapes is also discussed \cite{hooshyar2020variational}.

From the physical point of view, the complete self-similarity of the sediment-flux partitioning with $\mathcal{C_I}$ reveals that, even though fluvial erosion dominates over soil diffusion globally ($(2/\mathcal{C_I})\hat{S_b}\rightarrow 0$ and $\langle \hat{E}_{DL} \rangle \rightarrow 1$ in Eq. \eqref{glo_bud_nd_2}), the effect of soil diffusion does not fully disappear from the landscape dynamics. The resemblance of Fig. \ref{fig:one}A with the well-known Moody diagram for the turbulent friction coefficient, i.e., the proportion of kinetic energy loss due to viscous dissipation \cite{moody1944friction, munson2013}, presents an intriguing parallel with the self-similarity in wall-bounded turbulence in the complete turbulence regime with respect to the global Reynolds number. In both charts, the self-similar behavior arises for asymptotically large values of the control parameter ($\ln{\mathcal{C_I}} \gg 1$ \text{ and } $\ln{Re} \gg 1$) as diffusion becomes globally negligible compared to fluvial erosion (for landscapes) or advection (for fluid flows). Moreover, the vertical shift of the asymptotic $\hat{S_b}$ with the power-law exponent $m$ is reminiscent of shift due to the power-law exponent that defines the fluid rheology \cite{dodge1959turbulent, kawase1994friction} (see \ref{app:turbland} a detailed discussion).

Average landscape properties become invariant in the fluvial erosion regime. Fig. \ref{fig:one}D-E presents two landscapes from this self-similar regime, with $\mathcal{C_I}$ values differing by two orders of magnitude. Despite having different dimensional elevation fields ($z$ in green), these landscapes exhibit the same dimensionless relief (up to 0.75 as $\hat{z}$). Moreover, the normalized specific contributing area fields for these asymptotic solutions appear similar (Fig. \ref{fig:one}F), as do the exceedance-probability distributions with their power-law scaling (Fig. \ref{fig:one}G), highlighting the statistical self-similarity across different observation scales of each landscape geometry \cite{rodriguez2001fractal}. Incidentally, the obtained scaling exponent of $-0.45$ is close to the one observed in fluvial landscapes \cite{rigon1996hack, rinaldo2014evolution}.

\section{\label{S5}Diffusion localization at large $\mathcal{C_I}$}

The self-similar regime proceeds for several orders of magnitude of $\mathcal{C_I}$. Away from the ridgetops smoothed by diffusion, the valleys are mostly shaped by fluvial erosion. With increasing $\mathcal{C_I}$, the ridges become sharper, foreshadowing the development of singularities in the vanishing diffusion limit. While fluvial erosion dominates over diffusion globally at high $\mathcal{C_I}$, with $(2/\mathcal{C_I})\hat{S_b}\rightarrow 0$ and $\langle \hat{E}_{DL} \rangle \rightarrow 1$ in Eq. \eqref{glo_bud_nd_2}, the impact of diffusion persists, localized around these developing singularities.

\begin{figure*}[!bh]
\centering
\includegraphics[width=0.95\linewidth]{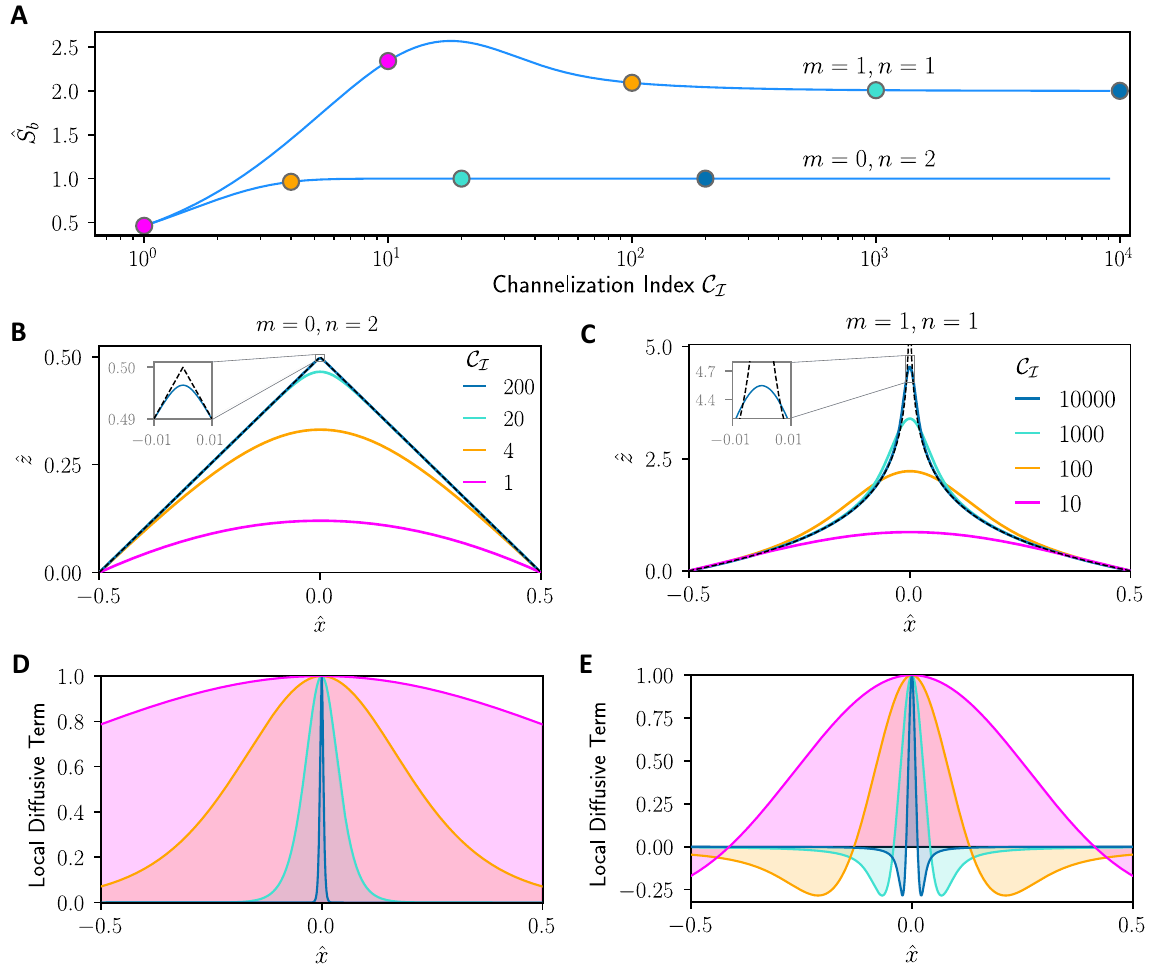}
\caption{\label{fig:two}Unchannelized 1D landscapes. (A) Global sediment budget chart ($\hat{S}_b$ vs $\mathcal{C_I}$) for $m=0,n=2$, and $m=n=1$ shows that the unchannelized landscapes also reach a self-similar plateau. (B and C) Steady-state landscapes for different values of $\mathcal{C_I}$. Sharp ridges develop as $\mathcal{C_I}$ increases for both cases. The dashed curves represent vanishing diffusion solutions. (D and E) Local sediment budget for the landscapes of B and C. The colored curves represent the local diffusion term. Unity on the $y$-axis represents the unit influx by uplift. At large $\mathcal{C_I}$, soil diffusion is crucial in removing the sediment influx on the landscape ridges.}
\end{figure*}

The vanishing diffusion (viscosity) limit has been studied in different problems of mathematical physics. In hydrodynamic turbulence, the convergence of vanishing viscosity solutions of the Navier-Stokes equation to weak solutions of (Euler) inviscid fluid equations has been shown for an unbounded domain \cite{kato1972nonstationary, swann1971convergence}. This convergence remains an open question for domains with boundary conditions, although it has been established for specific cases \cite{kato1984remarks, lopes2008vanishing, drivas2019remarks}. Zero viscosity limit in wave propagation problems leads to a sudden jump or shock discontinuity \cite{bec2007burgers, whitham2011linear, kluwick2018shock}. The theoretical emergence of singularities lies at the heart of admissibility issues; as noted by Earnshaw \cite{earnshaw1860viii}, singularity is obviously a physical impossibility, as `it is certain Nature has a way of avoiding its actual occurrence.' 

In mathematical analysis, the vanishing diffusion limit selects unique non-differentiable solutions of first-order fully nonlinear equations \cite{achdou2013hamilton}. Introduced by Crandall and Lions \cite{crandall1983viscosity, achdou2013hamilton} for Hamilton-Jacobi equations, the viscosity solution of the primary nonlinear equation is obtained through the convergence of the solution to a parabolic/elliptic formulation (achieved by introducing a Laplacian term) as the diffusivity/viscosity representing diffusion coefficient, heat conduction, fluid viscosity, etc., approaches zero. The convergent solution is a unique weak solution of the underlying non-linear equation.

In this Section, we show that soil diffusion provides the essential mechanism to obtain well-behaved topographies at large $\mathcal{C_I}$ values, remaining important where the landscape exhibits large concavity or convexity, i.e., in valleys and ridges. 

\subsection{\label{S51}Unchannelized 1D landscapes}

Although 1D solutions are less representative of natural landscapes due to the absence of channelization, they provide invaluable analytical insights. Two cases, $m=0$, $n=2$ and $m=n=1$, can be solved analytically (see \ref{app:ana}).

For $m=0,n=2$, the elevation equation corresponds to the well-known Kardar-Parisi-Zhang (KPZ) model for growing interfaces \cite{kardar1986dynamic}, with a constant uplift term, instead of the typical stochastic growth term. The elevation slope corresponds to the 1D Burgers' equation through the substitution $\hat{v} = -d\hat{z} /d\hat{x}$ \cite{bec2007burgers}. More generally, for $m=0$ and any real value of $n$, the elevation equation is the well-known eikonal equation employed in different branches of physics, engineering, and landscape modeling \cite{cornbleet1996eikonal, nechaev2017geometric, anand2023eikonal}. For $m=n=1$, fluvial erosion scales linearly with the specific contributing area and local slope, with the surface elevation acting as an active scalar that mutually interacts with the specific contributing area (discussed later).

\subsubsection{Effects of diffusion in the self-similar regime}

Fig. \ref{fig:two}A shows that the unchannelized 1D landscapes also exhibit complete self-similarity in the global sediment partitioning for large $\mathcal{C_I}$ values; this is akin to the self-similarity observed in channelized 2D landscapes (fluvial erosion regime in Fig. \ref{fig:one}A). Interestingly, this result highlights that the self-similar behavior of landscapes at asymptotically large $\mathcal{C_I}$ values is not contingent upon the presence of fractal drainage networks.

Fig. \ref{fig:two}B-C display the changes in the elevation profiles with increasing $\mathcal{C_I}$ (dashed curves in Fig. \ref{fig:two}B and C). At low $\mathcal{C_I}$, the landscapes exhibit near-parabolic elevation profiles for the two cases of erosion exponents. As $\mathcal{C_I}$ increases, a sharper ridgeline emerges in both scenarios, consistent with previous findings \cite{howard1997badland, tucker1998hillslope, whipple1999dynamics}. The elevation profile is linear for a major portion of the domain at large $\mathcal{C_I}$ values for $m=0$, with a sharp curvature near the ridgeline. While for $m=1$, the hillslopes are concave up moving from the boundary towards the ridge, resulting in the rapid growth of the slope and the sharpened (convex up) curvature confined to small scales near the central ridge.

Fig. \ref{fig:two}D-E displays the diffusion sediment flux term ($- \mathcal{C_I}^{-1} d^2 \hat{z}/d\hat{x}^2$) as a function of $\hat{x}$ for different $\mathcal{C_I}$ values. The uplift is represented by the horizontal line at unity, and therefore, the complementary height from unity to the diffusion curve represents the fluvial erosion at that point. Shaded regions below the colored curves show the local flux by soil diffusion, with positive/negative values indicating erosion/deposition. An increase in $\mathcal{C_I}$ corresponds to a reduction in the global sediment flux by diffusion, expressed by the integral of the shaded areas in Fig. \ref{fig:two}D-E. Nonetheless, diffusion remains vital, even at very large $\mathcal{C_I}$ values. In the case of $m=0$, diffusion does not contribute to the transport of sediments over most of the domain, but it consistently erodes around the ridge due to its sharp convexity. For $m=1$, diffusion erodes the area around the sharp convex ridgeline and deposits sediments close to the start of the sharp symmetric concave valleys. This localized mechanism plays a key role in stabilizing the topography as channelization index values become large. As the valley is predominantly shaped by fluvial erosion, the deposition of sediments by diffusion is negligible downstream where the contributing area increases.

\subsubsection{Singular ridges as shock waves for vanishing diffusion}

We now examine the 1D solutions as $D \rightarrow 0$ with boundary conditions $\hat{z}(\hat{x} \pm 0.5)=0$. In the case of $m=0,n=2$, the solution is the signed distance function (0.5 - $|\hat{x}|$) featuring a non-differentiable ridge at $\hat{x} = 0$, while with $m=n=1$, the solution consists of logarithmic hillslopes ($\ln{0.5} - \ln{|\hat{x}|}$) with a discontinuity at $\hat{x} = 0$. We can connect the ridge singularity in the zero diffusion case to the formation of shock waves. Since for $m=0, n=2$ we have a 1D Burgers' equation for the slope, nondifferentiability at ridgeline symbolizes an abrupt, yet finite change in $\hat{v}$ at $\hat{x} = 0$ (Fig. \ref{fig:v_SI}A). This result relates the singular ridge of 1D landscapes to a jump discontinuity in $\hat{v}$ ($= -d\hat{z} / d\hat{x}$) and the generation of shock waves in the inviscid Burgers' equation \cite{bec2007burgers, whitham2011linear}. For $m=n=1$, $\hat{v}$ has a vertical asymptote on both sides of $\hat{x} = 0$, representing an infinite discontinuity in $\hat{v}$ at $\hat{x} = 0$ (Fig. \ref{fig:v_SI}B).

These solutions can be seen as weak solutions when excluding soil diffusion from the surface evolution, a non-linear first-order partial differential equation referred to as the stream-power equation in geomorphology \cite{whipple1999dynamics, willett1999orogeny, Perron2009}. The stream-power equation parallels the inviscid Euler equations in fluid dynamics \cite{batchelor1967introduction}. According to the viscosity solution approach of Crandall and Lions \cite{crandall1983viscosity, achdou2013hamilton}, 1D vanishing diffusion solutions are unique viscosity solutions of the stream-power equation. The dashed curves in Fig. \ref{fig:two}B-C depict the vanishing diffusion solutions for both sets of exponents, along with the continuous curves for $\ln \mathcal{C_I} \gg 1$). While these solutions closely match over the majority of the domain, the contrast is apparent around the central ridgeline.

The different nature of singularity as a function of the fluvial law exponents in LEM bears resemblance to the distinct singularity types in wave turbulence \cite{kuznetsov2004turbulence}, where first order \cite{saffman1971spectrum, kadomtsev1973acoustic} and second order \cite{kuznetsov2004turbulence, ricard2023transition} wave discontinuities are observed for different geometries and degrees of anisotropy.

\begin{figure*}[!th]
\centering
\includegraphics[width=0.95\linewidth]{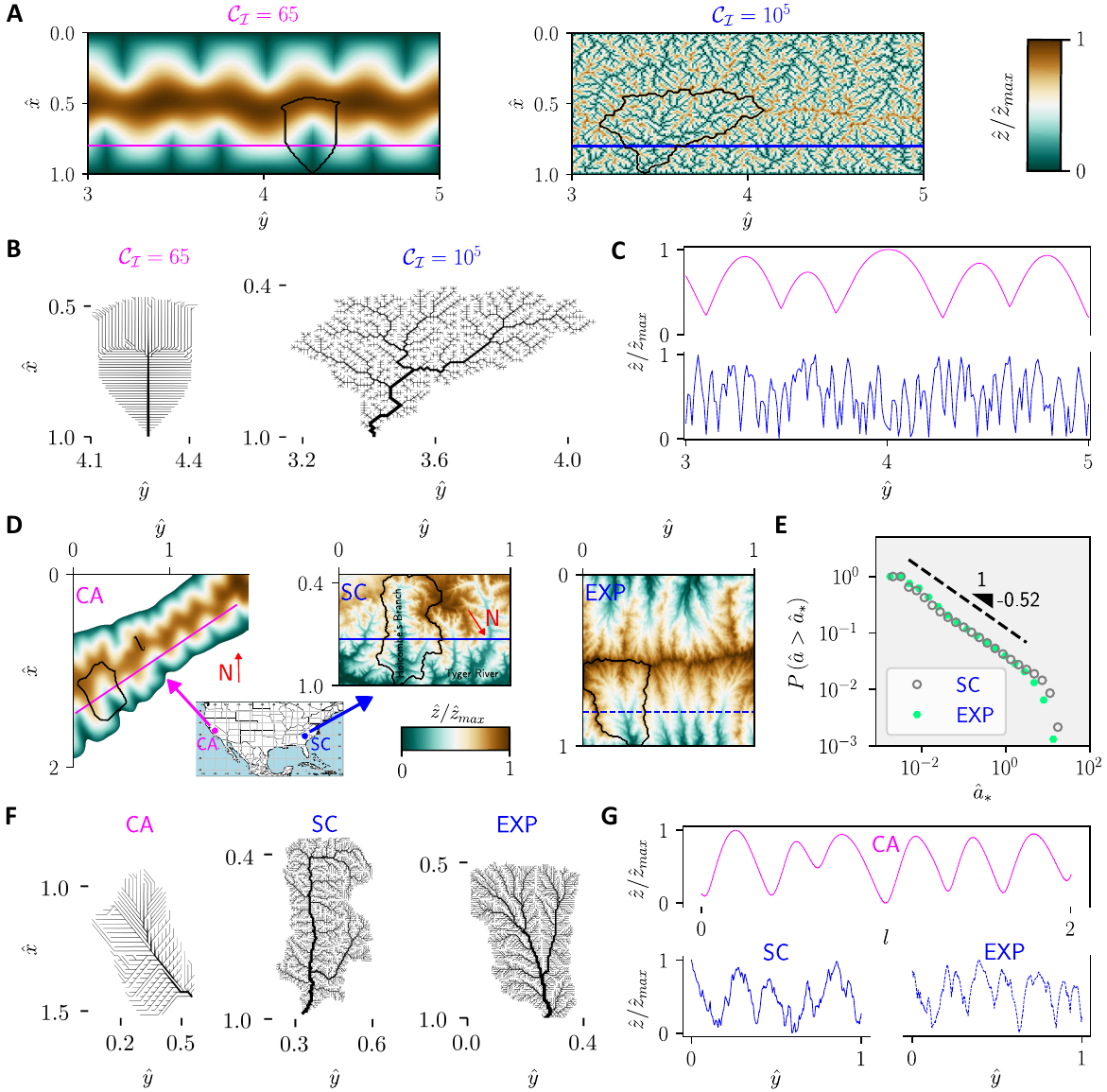}
\caption{\label{fig:three}Channelized 2D landscapes. (A-C) Numerical landscapes at increasing $\mathcal{C_I}$ values for $m=n=1$. (D-G) Natural (CA, SC) and experimental (EXP) topographies, ranging from equally spaced first-order valleys of CA ($\sim 1.1$ km${}^{2}$) to complex branching patterns in SC ($\sim 23$ km${}^{2}$) and EXP ($\sim 0.25$ m${}^{2}$). The estimated value of exponent $m$ is around $0.3$ in these cases. (A and D) Surface elevation fields. (B and F) The largest drainage basins, highlighted by the black curves in A and D. (C and G) 1D longitudinal transects along the colored lines of A and D show an increase in the number and curvature of convex ridges. (E) Exceedance probability distributions of $\hat{a}$ for SC and EXP show a remarkable collapse with a power-law exponent $\approx - 0.52$.}
\end{figure*}

\subsubsection{Analogy with Burgers vortex sheet}

1D solutions have also provided insight into self-similarity and localized diffusion effects in turbulent flows \cite{beronov1996linear, davidson2022incompressible}. Remarkably, with uplift varying linearly with elevation for $m=n=1$, the LEM equations correspond precisely to the well-known Burgers vortex sheet (see \ref{app:vortex}), the cartesian analog of the Burgers vortex tube \cite{burgers1948mathematical}), a well-known paradigm for fine-scale structures in incompressible turbulence \cite{townsend1951fine, moffatt1994stretched}. It depicts a 1D localized vortex layer that attains equilibrium through the hyperbolic stretching of a plain stagnation point flow, which is counteracted by the outward diffusion of vorticity \cite{sherman1990viscous, davidson2022incompressible}.

Analogously, the 1D steady elevation becomes a self-similar Gaussian profile for $U/K=1$, with no $\mathcal{C_I}$ dependence under the action of uplift, fluvial erosion, and diffusion, which matches exactly the steady self-similar Gaussian vortex sheet balancing advection, strain, and diffusion. As in the vortex sheet, which is characterized by a width scale formed by the kinematic viscosity and the imposed strain rate, here the landscape width is set by the scale $\sqrt{D/K}$, $D/K$ being the diffusion/fluvial coefficient, while the elevation gets more confined as $D/K \rightarrow 0$.

\subsection{\label{S52}Channelized 2D landscapes}

The localized diffusion effect at asymptotically large $\mathcal{C_I}$ is illustrated in 2D fluvial landscapes by analyzing numerical, natural, and experimental topographies.

We begin with 2D numerical simulations, shown in Fig. \ref{fig:three}A. The simulations depict two landscapes with varying degrees of surface elevation complexity. The first landscape exhibits a nearly periodic pattern of ridges and valleys ($\mathcal{C_I}=65$), while the second showcases a highly intricate network of ridges and valleys spanning multiple spatial scales ($\mathcal{C_I}=10^5$). The topographic complexity is apparent in the largest drainage basin of each landscape (Fig. \ref{fig:three}B) as well as in the elevation profiles along a 1D transect (Fig. \ref{fig:three}C). The elevation transects transition from a small number of smooth, nearly parabolic ridges separated by first-order valleys to a multitude of sharp convex ridges, primarily eroded by soil diffusion.

Similar to the numerical simulations, the natural and experimental landscapes of Fig. \ref{fig:three}D show varying levels of branching complexity, ranging from first-order valleys (CA) to heavily dissected topographies (SC, EXP). The Gabilan Mesa landscape in California (CA), covering an area of $\sim 1.06$ km${}^{2}$, features valleys spaced at roughly $163$ m intervals along a long ridge, with $m \approx 0.35$ \cite{Perron2009, anand2022inception}. The heavily dissected basin of the area around $23$ km${}^{2}$ draining to Tyger River in South Carolina (SC) has a much more complex topography and $m \approx 0.31$ \cite{hooshyar2021grl}. The channelized surface (EXP) comes from an experiment conducted on a $50 \text{ cm} \times 50 \text{ cm}$ sediment box (details in \cite{singh2015landscape}), with $m$ around $0.28$. 

Despite orders of magnitude differences in their spatial scales, a striking similarity can be observed between the complex natural (SC) and experimental (EXP) topographies. This similarity emerges in the elevation transects (Fig. \ref{fig:three}G), in the largest drainage basin (Fig. \ref{fig:three}F), and the remarkable collapse of exceedance probability distributions of the normalized specific contributing area fields (Fig. \ref{fig:three}E), with a scaling exponent around $-0.52$. This indicates that the same underlying physical mechanisms are at play, also noted previously in the 2D LEM solutions from the fluvial erosion regime (Fig. \ref{fig:one}D-G).

To highlight the localized effect of diffusion in numerical, natural, and experimental landscapes, we can focus on their longitudinal elevation transects (Fig. \ref{fig:three}C,G). At low $\mathcal{C_I}$, elevation fields (numerical -- $\mathcal{C_I} = 65$ and natural -- CA) are composed of alternate sequences of near-parabolic ridges separated by first-order valleys with concave curvature (pink curves). The ridges recall the unchannelized 1D landscapes dominated by soil diffusion. At increased fluvial erosion intensity, i.e., large $\mathcal{C_I}$, the channelization cascade leads to the emergence of small-scale fluctuations in the elevation fields (numerical -- $\mathcal{C_I} = 10^5$, natural -- SC, and experimental -- EXP), which become an increasingly complicated superimposition of long- and short-frequency modes (blue curves). A higher number of ridges develop, with sharp convex curvatures stabilized by diffusion following the same mechanism as shown in Fig. \ref{fig:two}D-E. 

\section{\label{S6}Discussion}

\subsection{Fluvial landscapes and turbulent flows}

The numerous analogies between landscapes and turbulence suggest a similar nonlinear cascade mechanism, operating at increasingly smaller scales, giving rise to self-similar asymptotic behavior in both systems. Following landscapes upstream, one finds a bifurcation to smaller and smaller channels, akin to the energy cascade of vortices in turbulence \cite{bonetti2020channelization, porporato2022hydrology}. In the self-similar regimes, local properties of the landscape elevation and the turbulent velocity fields share similar features in the logarithmic profiles \cite{katul2019primer, hooshyar2020pre} and the power spectra \cite{frisch1995turbulence, hooshyar2021grl}.

Our results further strengthen this analogy by showing that the average sediment-flux partitioning over the entire landscape domain exhibits self-similar behavior for large $\mathcal{C_I}$ values, analogous to the self-similar energy partitioning of bounded turbulent flow at large $Re$ values \cite{moody1944friction, munson2013}. LEM solutions, supported by real-world observations, reveal that under intense fluvial erosion, diffusion remains crucial to prevent the formation of singularities at sharp ridges, where fluvial erosion cannot occur due to the contributing area tending towards zero. This localized soil diffusion action in landscapes is comparable to the viscous dissipation in the turbulent energy cascade, where kinetic energy is converted into heat energy at small (Kolmogorov) scales \cite{batchelor1967introduction, frisch1995turbulence}. The analogy with the Burgers vortex sheet presented earlier provides an exact parallel, illustrating the balance between localized diffusion and erosion/vortex stretching \cite{burgers1948mathematical, townsend1951fine}.

\subsection{Surface elevation as active scalar}

The surface elevation operates as an active scalar that mutually interacts with the specific contributing area \cite{friedlander2021vanishing}. For $m=n=1$, the fluvial erosion term can be written as $\Vec{q}(\hat{z}) \cdot \hat{\nabla} \hat{z}$, where $\Vec{q}=-\hat{a}\hat{\nabla} \hat{z}/|\hat{\nabla} \hat{z}|$. This formulation denotes the advective term of an active scalar, namely $\Vec{u}(\theta) \cdot \nabla \theta$, $\theta$ being the active scalar that mutually interacts with the velocity field $\Vec{u}$. While $\Vec{u}$ is usually a divergent-free vector in fluid dynamics, $\Vec{q}$ is the flux based on the rainfall input in LEMs and it typically aligns with the topographic gradient \cite{bonetti2018theory, anand2022inception}. Similar equations appear in geophysical fluid dynamics \cite{moffatt1994magnetostrophic, held1995surface}, and understanding how singularities dissipate and give rise to sharp solutions in the vanishing diffusion limit is at the forefront of research in these fields.

\subsection{Linkage to optimal channel networks}

A previous study \cite{hooshyar2020variational} emphasized the connection between the optimality principle of the LEM in the continuous domain under negligible soil diffusion globally and the optimal channel network (OCN) theory for discrete drainage network configurations \cite{rinaldo1993self, rodriguez2001fractal, balister2018river}. In OCN simulations, the exceedance probability distribution of the contributing area exhibits a power-law exponent of around -0.43 in sub-optimal feasible networks \cite{rodriguez2001fractal, rinaldo2014evolution, balister2018river}. These networks are obtained iteratively by minimizing energy dissipation, starting from a random configuration. The narrow range of natural catchments' power-law exponents between -0.42 and -0.45 \cite{rinaldo2014evolution} suggests that sub-optimal OCNs adequately capture the statistical character of drainage networks, which approach a feasible state over long geological timescales.

LEM solutions in the fluvial erosion regime also display the same statistical behavior. Fig. \ref{fig:one}G shows that the power-law exponent of $P \left(\hat{a} > \hat{a}_* \right)$ for two landscapes in the self-similar regime for $m=0.5$ is around -0.45, matching the signature of feasible optimality in OCNs and natural drainage networks. This result provides compelling evidence that not only do the steady LEM solutions exhibit self-similar behavior for very large $\mathcal{C_I}$, but they also have the scaling signature associated with sub-optimal landscapes obtained as OCNs in discrete lattice geometries with no explicit soil diffusion. Presumably, the grid size sets the diffusive scale and the related channelization index value in OCN.

\subsection{Coexistence of complete and incomplete self-similarity}

The connection between the fluvial erosion regime of LEM and the OCN statistics indicates the coexistence of two types of self-similarity. On one hand, complete self-similarity is observed at large $\mathcal{C_I}$ values for quantities related to mean sediment fluxes and elevation; on the other hand, self-similarity is also observed in the statistical features of each landscape geometry. The latter is typically linked to the fractal properties of landscapes \cite{rodriguez2001fractal}, which, due to their irrational scaling exponents, have been associated with incomplete (or second type) self-similarity \cite{barenblatt1996scaling}.

A similar situation is found in fluid turbulence, where the logarithmic law for the velocity profile, the local spectral structure of fully developed turbulent flows, and the fully rough regime in the Moody diagram point to complete self-similarity at very high Reynolds numbers \cite{barenblatt1996scaling, katul2019primer}. At the same time, the statistically self-similar flow features and fractal geometries of fully developed turbulent flows suggest incomplete self-similarity \cite{sreenivasan1986fractal, heisel2022self, barenblatt1995does}.

The simple example of the Kock triad can shed light on the spectrum of self-similar behaviors observed in landscapes and turbulent flows. With an increase in the number of sides, the perimeter of the Kock curve approaches infinity with a fractal dimension $\approx 1.26$, which is an indicator of incomplete self-similarity \cite{barenblatt1996scaling}. However, the enclosed area converges to 8/5 times the original area as the number of sides becomes very large, which is indicative of complete self-similarity. This simple geometric model shows that complex physical problems can exhibit different types of self-similar behavior; the specific form of self-similarity that emerges in a problem depends on the attributes of the solutions considered.

\section{Conclusion}

Landscape classification into three regimes (Fig. \ref{fig:one}A) shows that, just as laminar flow is the exception rather than the rule, so are smooth landscapes without valleys from the diffusion regime. The most frequent topographies are completely self-similar with respect to the channelization index, while also exhibiting fractal geometries with incomplete self-similarity.

In the limit of infinite channelization indices, in the far right of the complete self-similarity plateau, as fluvial erosion intensity becomes dominant over diffusion processes everywhere except the network of ridges (and possibly valleys), the topographic singularities become analogous to discontinuities in surface growth models and to the shock generation in wave propagation. In all these systems, diffusion, however small, remains in place to regularize singularities, much like how `momentum must ultimately pass from the eddies to the ground by means of the almost infinitesimal viscosity of air,' Taylor \cite{taylor1915eddy}.

\section*{Data \& code availability} 
High-resolution elevation data for natural topographies can be obtained from \url{https://opentopography.org}. The details of the numerical solver are described in \cite{anand2020ems} and the Python code is available on \url{https://github.com/ShashankAnand1996/LEM} \cite{anand_2022_6824822}.

\section*{Acknowledgements}
\label{S7}
The authors acknowledge support from the BP Carbon Mitigation Initiative at Princeton University and the High Meadows Environmental Institute. S.K.A. acknowledges the generous support of the Thomas Perkins Class of 1894 Graduate Fellowship Fund. The research of T.D. was partially supported by NSF--DMS grant 2106233 and an NSF CAREER award \#2235395. The numerical simulations were performed on computational resources provided by Princeton Research Computing at Princeton University. 

High-resolution elevation data for natural topographies can be obtained from \url{https://opentopography.org}. The details of the numerical solver are described in \cite{anand2020ems} and the Python code is available on \url{https://github.com/ShashankAnand1996/LEM} \cite{anand_2022_6824822}.

\bibliographystyle{elsarticle-num}

\bibliography{reference}
\appendix
\section{Interpreting the Channelization Index\label{app:CI}}

The model has three primary dimensions: horizontal, vertical, and time. Using $K$, $U$ and $l$ as scaling variables, the following non-dimensional quantities are obtained: $\hat{z} = z \left(K l^{m-n} /U \right)^{1/n}$, $\hat{a} = a/l$, $\hat{t} = t \left(K l^{m-n} / U^{1-n} \right)^{1/n}$, $\hat{x}=x/l$, and $\hat{y}=y/l$ to get dimensionless Eqs. \eqref{eq:z_nd} and \eqref{eq:a_nd}. The dimensional analysis yields one control parameter, the channelization index $\mathcal{C_I}$ \cite{bonetti2020channelization}.

Just as the Reynolds Number $Re$, derived as a ratio of inertial and diffusion term in the Navier-Stokes equations, can be adapted to describe the ratio of diffusion to inertial time-scales, turbulent to molecular viscosity \cite{tennekes2018first}, $\mathcal{C_I}$ can be interpreted in multiple ways. $\mathcal{C_I}$ describes the relative control of fluvial erosion to soil diffusion in the uplifted topography. It has also been written as dimensionless boundary size for the parameterless governing equations \cite{anandcomment2022, replycommentlitwin}. $\mathcal{C_I}$ can also be interpreted as the ratio of the two time scales as
\begin{equation}
    \mathcal{C_I} = \frac{l^2/D}{U^{1/n-1}/\left( K l^{m-n}\right)^{1/n}} = \frac{\text{Soil diffusion time scale}}{\text{Fluvial erosion time scale}}. 
\end{equation}
This understanding of $\mathcal{C_I}$ provides an intriguing linkage of the LEM solutions with the two-phase evolution theory for fluvial-dominated landscapes \cite{banavar2001scaling}, where fluvial erosion acts at a relatively fast time scale (freezing time) at high $\mathcal{C_I}$ to practically freeze the 2D planar ridge/valley organization. While diffusion smooths high-frequency modes of the 3D elevation field over a much longer duration (relaxation time) until the steady state is reached. 

\section{Global Sediment Budget \label{app:gsb}}

For a domain $\Omega$ covering a total area $\mathcal{A}$, the steady-state sediment budget is written as
\begin{equation}
\label{app_glo_bud_nd}
        \int_\Omega \frac{1}{\mathcal{A}}\left(\frac{1}{\mathcal{C_I}}\hat{\nabla}^2 \hat{z} - \hat{a}^m |\hat{\nabla} \hat{z}|^n + 1 \right) d\Omega = 0.
\end{equation}

We simplify the average diffusion flux flowing out of the boundaries using the divergence theorem as
\begin{equation}
\label{app_diff_bud_nd}
\frac{1}{\mathcal{A}} \int_\Omega \left(\frac{1}{\mathcal{C_I}}\hat{\nabla}^2 \hat{z} \right) d\Omega = \frac{1}{\mathcal{A} \mathcal{C_I}} \oint \hat{\nabla} \hat{z} \cdot \vec{n} d \mathcal{B} = - \frac{\hat{S_b}}{\mathcal{C_I}} \frac{l_\mathcal{B}}{\mathcal{A}},
\end{equation}
where $\vec{n}$ is the normal vector to the boundary ($\mathcal{B}$), $\hat{S_b}$ is the average boundary slope, and $l_\mathcal{B}$ is the boundary length. For a semi-infinite domain of unit width, $l_\mathcal{B}/\mathcal{A}$ limits to $2$.

We simply write the fluvial term as $\langle \hat{E}_{DL} \rangle$, and the uplift term is unity. Substituting these in Eq. \eqref{app_glo_bud_nd}, global sediment budget reads
\begin{equation}
\label{app_glo_bud_nd_2}
        \frac{2}{\mathcal{C_I}} \hat{S_b} + \langle \hat{E}_{DL} \rangle= 1.
\end{equation}

\section{Analytical solutions\label{app:ana}}

To obtain analytical expressions for 1D steady-state solutions as well as the variation of $\hat{S_b}$ as a function of $\mathcal{C_I}$, we consider unchannelized landscapes consisting of a single ridge in the center and a symmetric elevation profile that decreases monotonically on either side of it. Without loss of generality, $\hat{x}=0$ is fixed at the ridgeline that results in $\hat{a}= |\hat{x}|$. Eq. \eqref{eq:z_nd} at steady-state reads
\begin{align}
\frac{1}{\mathcal{C_I}} \frac{d^2\hat{z}}{d\hat{x}^2} - |\hat{x}|^m \left|\frac{d \hat{z}}{d\hat{x}}\right|^n + 1 = 0.\label{lem_eq_nd1d}
\end{align}

With zero elevation boundary conditions $\hat{z}\left( \hat{x} = \pm 1/2 \right) = 0$, analytical solution for $m=0$ and $n=2$ reads
\begin{align}
\label{lem_solm0n2_nd1d}
\hat{z}\left(\hat{x}\right) = \hat{x}-\frac{1}{2}-\frac{1}{\mathcal{C_I}} \log \left(\frac{\exp\left(2\,\mathcal{C_I}\,\hat{x}\right)+1}{\exp\left(2\,\mathcal{C_I}\right)+1}\right).
\end{align}

Differentiating Eq. \eqref{lem_solm0n2_nd1d} and substituting $\hat{x}=-0.5$, we get an analytical expression for the boundary slope as
\begin{align}
\label{sb_solm0n2_nd1d}
\hat{S_b}\left(\mathcal{C_I}\right) = \frac{\mathrm{e}^{\mathcal{C_I}}-1}{\mathrm{e}^{\mathcal{C_I}}+1}.
\end{align}

For $m=n=1$, the elevation field \cite{bonetti2020channelization} is
\begin{align}
\label{lem_solmn1_nd1d}
\hat{z}\left(\hat{x}\right) = \frac{\mathcal{C_I}}{2} \left[ \frac{1}{4} { }_{p}F_q \left(1,1;\frac{3}{2},2;-\frac{\mathcal{C_I}}{8}\right) - \hat{x}^2 { }_{p}F_q \left(1,1;\frac{3}{2},2;-\frac{\mathcal{C_I} \hat{x}^2}{2}\right) \right],
\end{align}
where ${ }_{p}F_q(., .; ., .; .)$ is the generalized hypergeometric function. Further, $\hat{S_b}$ is derived as a function of $\mathcal{C_I}$
\begin{align}
\label{sb_solmn1_nd1d}
\hat{S_b}\left(\mathcal{C_I}\right) = \sqrt{2\mathcal{C_I}} \mathcal{D}\left(\sqrt{\frac{\mathcal{C_I}}{8}}\right),
\end{align}
where $\mathcal{D}\left(\cdot\right)$ is the Dawson function.

In Fig. \ref{fig:two}A, Eqs. \eqref{sb_solm0n2_nd1d} and \eqref{sb_solmn1_nd1d} display the variation of $\hat{S_b}$ with increasing values of $\mathcal{C_I}$ for $m=0, n=2$ and $m=n=1$, respectively. Fig. \ref{fig:two}B-C show the analytical elevation profiles for both cases (Eqs. \eqref{lem_solm0n2_nd1d} and \eqref{lem_solmn1_nd1d}) at different $\mathcal{C_I}$ values.

\section{Burgers vortex sheet\label{app:vortex}}
We compute the 1D steady-state solution of Eqs. \eqref{eq:z_nd} and \eqref{eq:a_nd}, for which uplift varies linearly with the elevation field ($Uz$, instead of $U$ in the original formulation). With the horizontal length scale defined as $\sqrt{D/K}$ (instead of imposing an external length scale) and the time scale as $1/K$, the dimensionless equation reads
\begin{align}
\label{eq:lem_burg}
\frac{d^2\hat{z}}{d\hat{x}^2} + \hat{x} \frac{d \hat{z}}{d\hat{x}} + \frac{U}{K}\hat{z} = 0.
\end{align}
The steady profile for $\hat{z} = \hat{z}_\text{o}$ at $\hat{x} = 0$ and $z \rightarrow 0$ as $\hat{x} \rightarrow \pm \infty$ is
\begin{align}
\label{eq:z_1f1}
\hat{z}\left(\hat{x}\right) = \hat{z}_\text{o} { }_{1}F_1 \left(\frac{U}{2K},\frac{1}{2};-\frac{\hat{x}^2}{2}\right),
\end{align}
where ${ }_{1}F_1 \left(\cdot\right)$ is the Kummer confluent hypergeometric function.

For $U/K = 1$, Eq. \eqref{eq:lem_burg} is the steady-state vorticity equation for the Burgers vortex sheet \cite{sherman1990viscous}, where the horizontal scale is defined as the $\sqrt{\nu/\alpha}$ ($\nu$ is the kinematic viscosity and $\alpha$ is strain rate). In this case, Eq. \eqref{eq:z_1f1} gets simplified as a self-similar Gaussian profile for the elevation/vorticity as  
\begin{align}
\label{eq:z_gau}
\hat{z}\left(\hat{x}\right) = \hat{z}_\text{o} \exp{(-\hat{x}^2/2)}.
\end{align}

\begin{figure}[!htb]
\centering
\includegraphics[width=\linewidth]{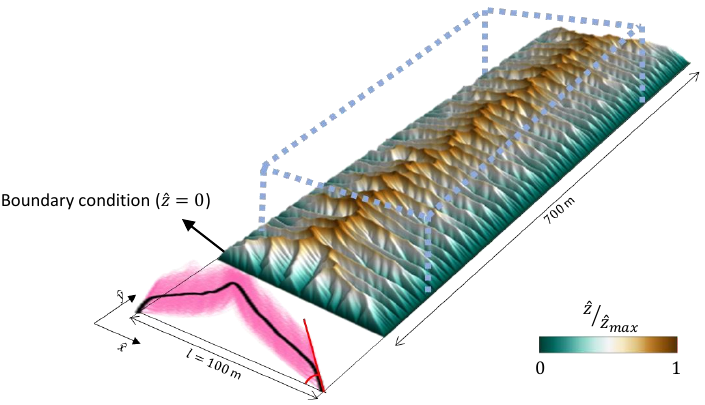}
\caption{\label{fig:numerical_setup}Model set-up used in this study. The example of a steady-state elevation field is shown for $\mathcal{C_I}=10^3$ with $m=0.5$ and $n=1.0$ in a rectangular domain (width = 100 m, length = 700 m). The dashed region represents the spatial range of the solution considered in the analysis that excludes the last 100 m from both sides along the length. This is done to reduce the effect of lateral fixed-elevation boundary conditions. Pink curves show the ensemble of the elevation along the length. The black curve is the mean elevation profile obtained from the transects' ensemble and the red line indicates the average boundary slope, which was computed as the slope of the linear fit to the mean profile in the first 3 m from the boundary.}
\end{figure}

\section{Simulation set-up\label{app:simset}}

We used here the numerical algorithm developed in \cite{anand2020ems} for efficient LEM simulations. The algorithm belongs to task-scheduling problems, where the flow network draining the landscape is traversed to provide efficient and accurate solutions. The algorithm has been thoroughly verified for mean elevation dynamics of transient solutions and the accuracy of steady-state solutions using a spatial convergence test \cite{anand2020ems}. The numerical results were recently validated against the onset of channelization, demonstrating good agreement with the linear stability analysis of analytical solutions for different $m$ and $n$ values \cite{anand2022inception}. 

All the simulations shown in this study were performed on a 700 m long and 100 m wide rectangular domain (unit grid size) with zero elevation boundary conditions. To reduce the impact of lateral boundary conditions and approximate a semi-infinite domain numerically (see Fig. \ref{fig:numerical_setup}), the last 100 m from the far ends were neglected during the analysis of the numerical solutions.

\begin{figure}[!ht]
\centering
\includegraphics[width=\linewidth]{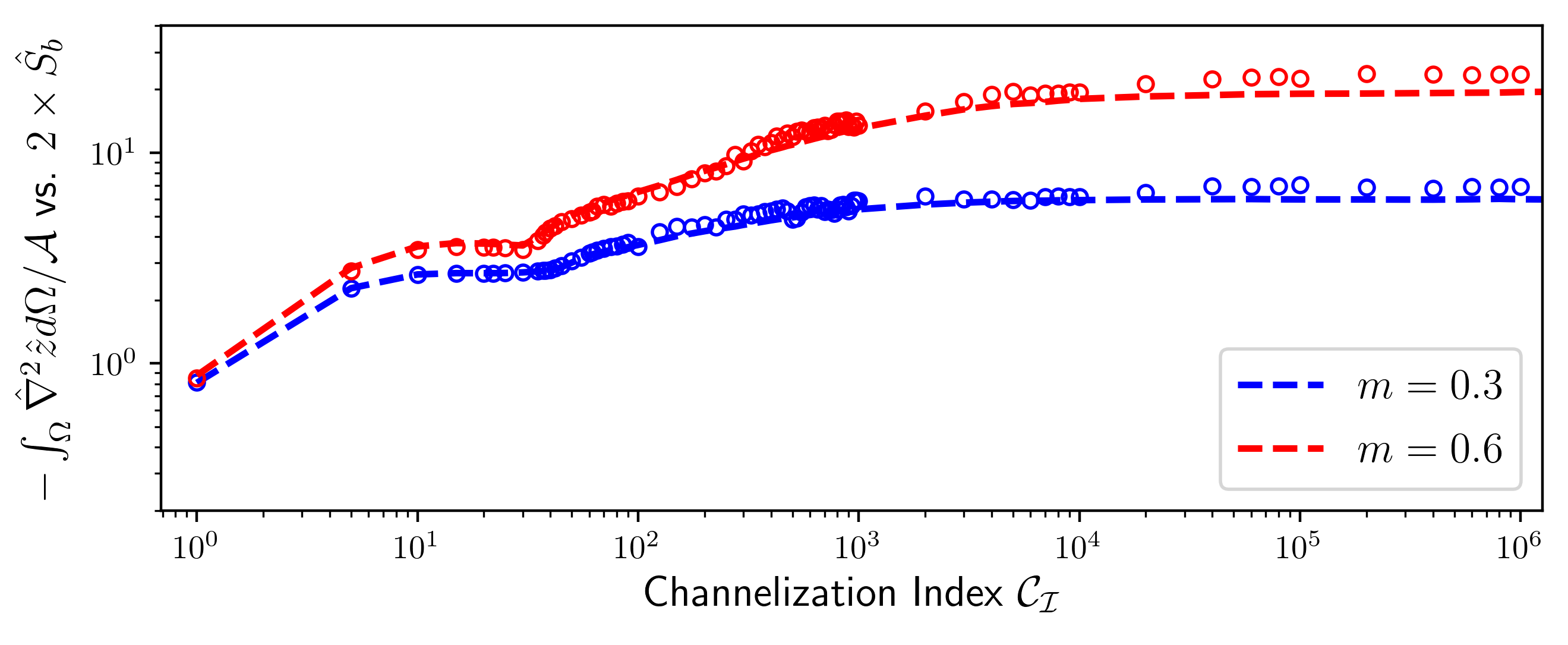}
\caption{\label{fig:laplace_comp}Comparison of the integral of soil diffusion (circles) with average boundary slope (dashed curves) for exponent $m=0.3$ (blue) and  $m=0.6$ (red). The relative error in both calculation methods is less than 13\% in 160 different $\mathcal{C_I}$ simulations shown here, with outgoing diffusion flux reaching a limiting value at high channelization index $\mathcal{C_I}$.}
\end{figure}

\section{Verification tests\label{app:vertest}}

The average boundary slope considers the total diffusion flux coming out of the domain. Using the divergence theorem, the integration of the diffusion term over the entire domain area $\mathcal{A}$ can be simplified as
\begin{equation}
\label{si_diff_bud_nd}
\frac{1}{\mathcal{A}} \int_\Omega \left(\frac{1}{\mathcal{C_I}}\hat{\nabla}^2 \hat{z} \right) d\Omega = - \frac{\hat{S_b}}{\mathcal{C_I}} \frac{l_\mathcal{B}}{\mathcal{A}},
\end{equation}
where $\hat{S_b}$ is the average boundary slope of the elevation field and $l_\mathcal{B}$ is the boundary length enclosing the region of interest. The negative diffusion flux averaged over the domain shows that overall, soil diffusion erodes more from convex ridges than it deposits in concave valleys. For a long piece of land with parallel boundaries considered here, $l_\mathcal{B}/\mathcal{A}$  limits to 2. Fig. \ref{fig:laplace_comp} provides a numerical validation of Eq. \eqref{si_diff_bud_nd} by comparing the diffusion integration over the entire domain with $\hat{S_b}$ for two different values of exponent $m =0.3$ and $m=0.6$ in a long rectangular domain (100 m wide and 700 m long) and $n=1$. The relative difference is less than 13\% in all 160 different $\mathcal{C_I}$ simulations.

\begin{figure}[!hbt]
\centering
\includegraphics[width=\linewidth]{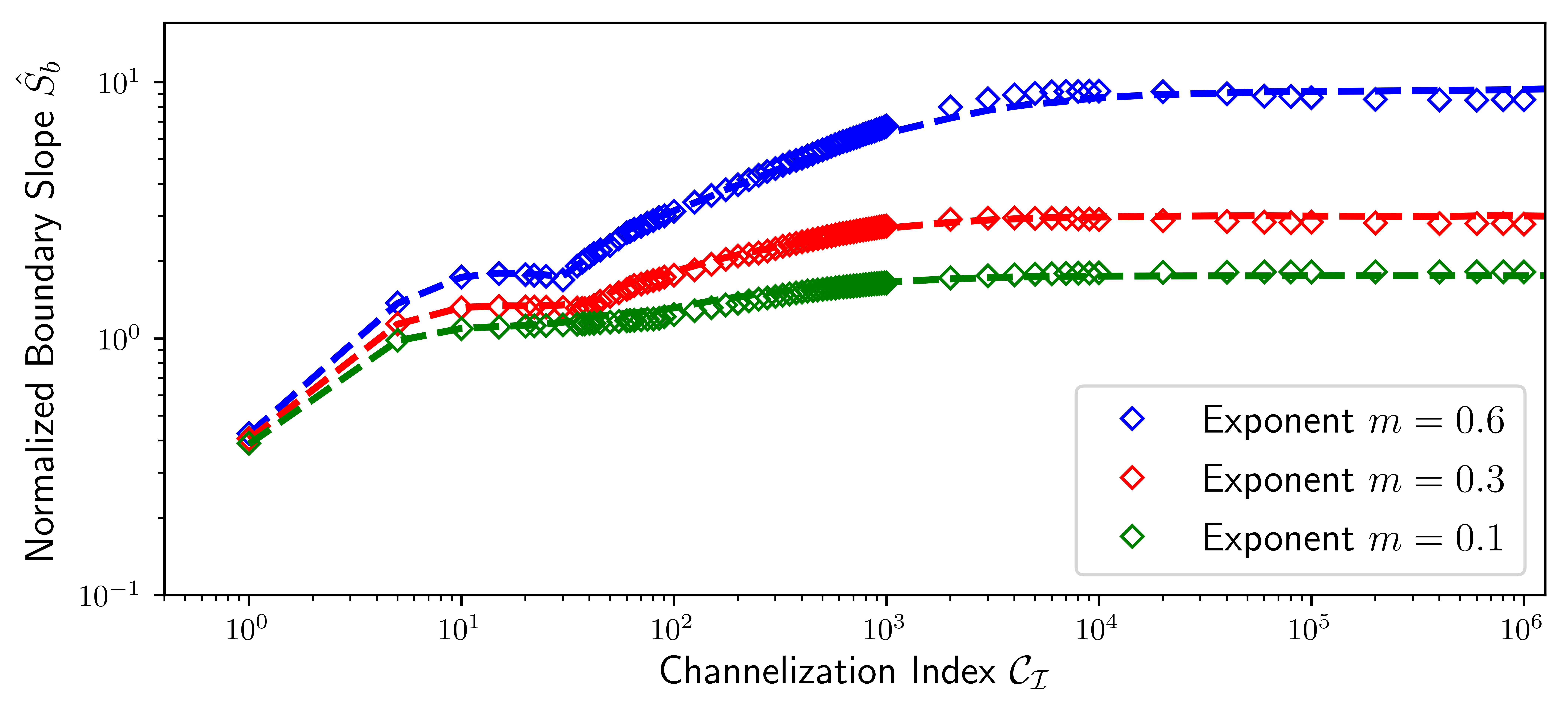}
\caption{\label{fig:resol_comp}Effect of grid resolution on the sediment partitioning results of numerical simulations. Comparison of the average boundary slope at resolution $1$ m (dashed curves) and $0.5$ m (diamond symbols) for exponent $m=0.1$ (green), $m=0.3$ (red), and $m=0.6$ (blue). The relative error in the computed average boundary slope is less than $4\%$ for $m=0.1$, less than $3\%$ for $m=0.3$, and less than $9\%$ for $m=0.6$.}
\end{figure}

To further verify that the presented results do not suffer from detrimental grid discretization effects, we performed numerical simulations at a resolution $dx=dy=0.5$ m for three different values of $m=0.1$, $m=0.3$, and $m=0.6$ and compared them with the results for $dx=dy=1.0$ m. Fig. \ref{fig:resol_comp} shows the compiled results of a good agreement of average boundary slope in all three exponent values. We also report that the relative error in computing $\hat{S_b}$ for two different resolution cases is less than $4\%$ for $m=0.1$, $3\%$ for $m=0.3$, and $9\%$ for $m=0.6$.

\section{Similarity of dimensional functions in bounded turbulence and landscapes\label{app:turbland}}

We consider the fluid flow in a long circular pipe for the power-law (PL) shear stress model  \cite{dodge1959turbulent, kawase1994friction}
\begin{equation}
    \tau = k_\tau\left(-\frac{du}{dr}\right)^{m_\tau},
\end{equation}
where $k_\tau$, the proportionality constant, becomes the dynamic viscosity ($\mu$) for the Newtonian fluid ($m_\tau=1$).

Similar to the dimensional analysis for the sediment budget of the LEM, the physical law for the wall shear stress is written as
 \begin{equation}
     \tau_w = \Phi_1 \left( v_c, d, \rho; k_\tau, m_\tau \right),
 \end{equation}
where $v_c$ is the characteristic velocity of the flow (generally taken as the mean velocity $V_m$), $d$ is the diameter of the circular pipe, and $\rho$ is the fluid density. Selecting $v_c$, $d$, and $\rho$ as the repeating variables, $\Pi$ theorem can be used to write
\begin{equation}
     \frac{\tau_w}{\rho v_c^2}= \Phi_2 \left( \frac{k_\tau}{\rho d^{m_\tau} v_c^{2-{m_\tau}}}, m_\tau \right).
\end{equation}

With some rearrangements and using $v_c =V_m$, the Darcy friction factor $f_D$ is obtained as
\begin{equation}
\label{eq:fd}
     {f_D} = \frac{\tau_w}{\frac{1}{2}\rho V^2} = \Phi_3 \left( {Re}_{PL}, m_\tau \right)
\end{equation}
where ${Re}_{PL} = \frac{\rho d^{m_\tau} V_m^{2-{m_\tau}}}{k_\tau}$ is a generalized Reynolds number for fluids with different rheology \cite{dodge1959turbulent} and ${f_D}$ is the generalized friction factor, characterizing the viscous dissipation per unit kinetic energy for a unit pipe length.

Eq. \eqref{eq:fd} here is analogous to Eq. \eqref{eq:phi_1} for the sediment partitioning based on $\hat{S_b}$ (fixed $n=1$) as
\begin{equation}
\label{f_S_b_1}
    \hat{S_b} = \varphi_1 \left( \mathcal{C_I}, m \right).
\end{equation}

\begin{figure}[!hbt]
\centering
\includegraphics[width=0.9\linewidth]{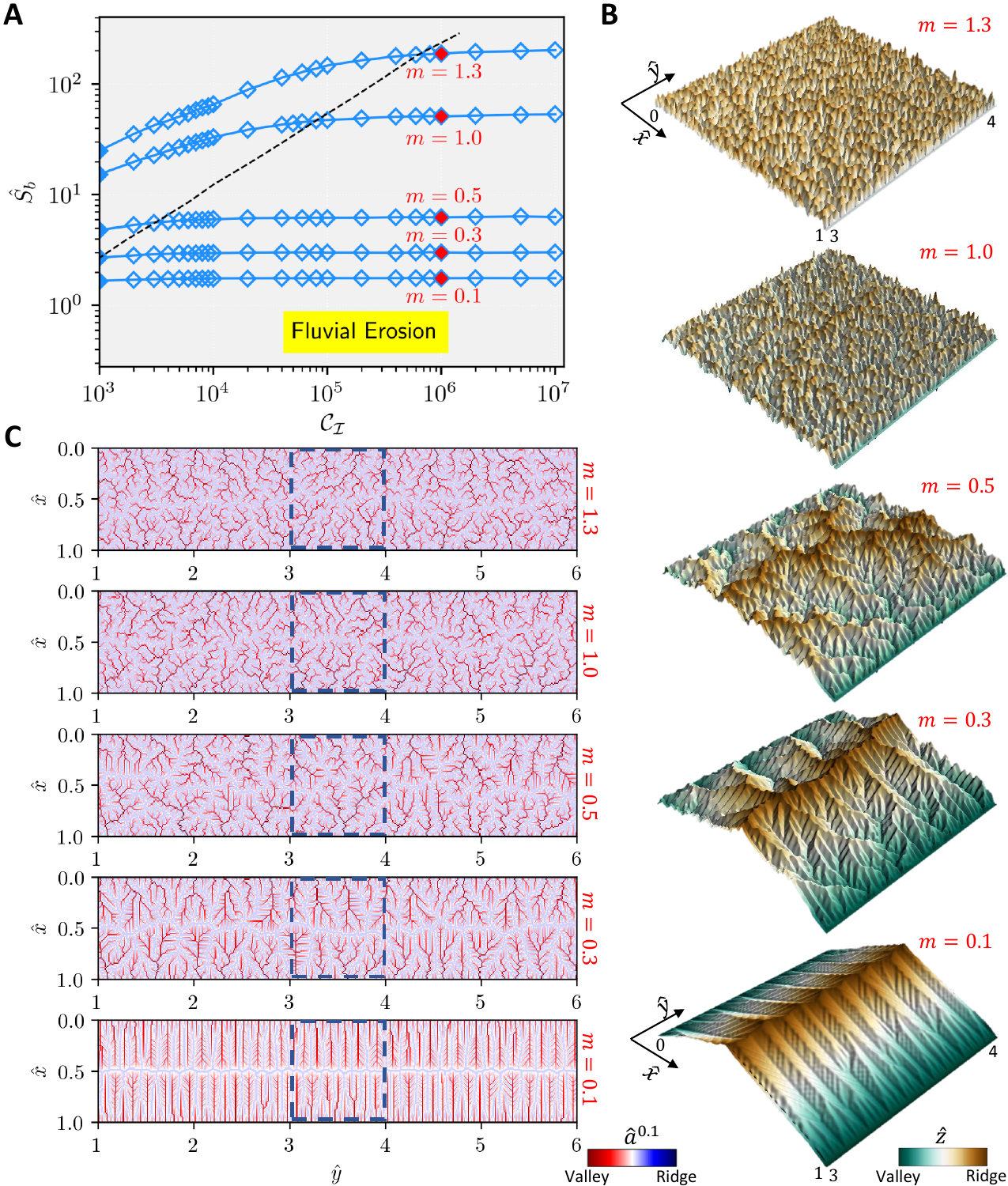}
\caption{\label{fig:exp_m_effect}Effect of $m$ on the self-similar landscape organization in fluvial erosion regime. (A) Inset of the sediment budget chart from Fig. \ref{fig:one}A. Red symbols mark landscapes obtained at $\mathcal{C_I}=10^6$ for different values of exponent $m$ and $n=1$. (B) 3D surface plots of the elevation fields and (C) specific contributing area $a$ fields for the selected landscapes. Dashed squares denote the regions of the plotted elevation fields in B. With the increased value of the exponent $m$, the landscapes lose the sense of direction with the development of a channelization cascade.}
\end{figure}

This result shows the dimensional functions' similarity in both bounded turbulent flows and fluvial landscapes, with $m$ playing the part of rheology in the evolving landscape for $\ln{\mathcal{C_I}} \gg 1$.

\section{Influence of the erosion exponents on the self-similar regime\label{app:diffmn}}

We explore here the influence of the erosion exponents $m$ and $n$ on the landscape self-similar regime. First, we keep $n$ fixed to unity and highlight how the exponent $m$ largely affects the sediment-flux partitioning and the steady-state landscape morphology (Fig. \ref{fig:exp_m_effect}).

The landscape exhibits a regular pattern of valleys and ridges for low $m$ values (e.g., $m=0.1$), featuring a prominent central ridge that divides the domain into two nearly symmetric parts \cite{shelef2014symmetry}. A large-scale flow directionality from the ridges to the valleys is observed. As $m$ increases, the central ridge becomes less regular and expands, encompassing a more significant portion of the domain ($m=0.3, 0.5$). With increasing $m$, the directionality of the elevation field becomes less apparent, which corroborates findings from previous studies \cite{howard1994detachment, tucker2002topographic, shelef2018channel}. For even larger $m$ values, the directionality of the ridge-valley network is completely lost ($m=1,1.3$). The landscape morphologies at such large $m$ values are very complicated, with a cascade of ridges and valleys from the large to the small scales. This cascade results in a bigger portion of the landscape domain occupied by ridges. Because soil diffusion erodes primarily on ridges, the expansion of the ridge network with growing values of $m$ entails that overall diffusion erodes more at higher $m$ values in the fluvial erosion regime, i.e., $\hat{S_b}$ is higher.

\section{Variational principle for asymptotically large channelization indices\label{app:varp}}

\begin{figure}[!hbt]
\centering
\includegraphics[width=\linewidth]{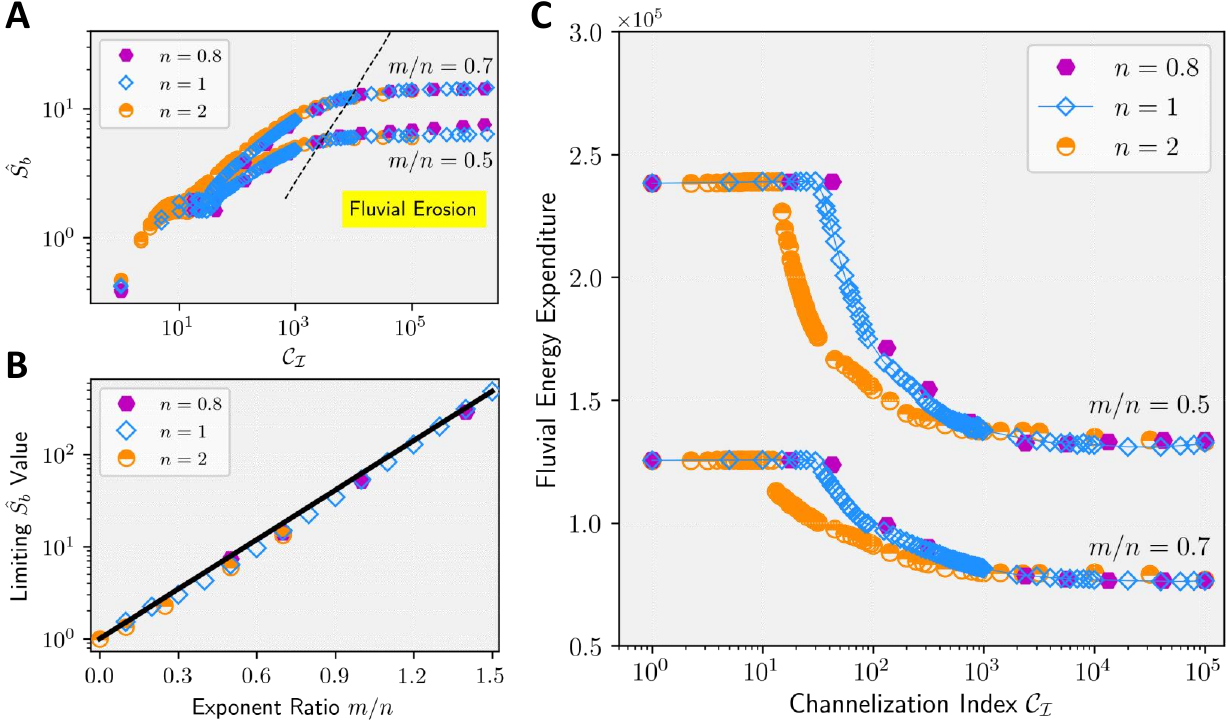}
\caption{\label{fig:m_n_ratio}The ratio $m/n$ characterizes the fluvial erosion regime. (A) $\hat{S}_b$ as a function of $\mathcal{C_I}$ for different values of $n$ (0.8, 1, 2) and $m$ that keep the $m/n$ ratio equal to 0.5 or 0.7. (B) Limiting $\hat{S}_b$ values in the self-similar regime at large $\mathcal{C_I}$ as a function of the ratio $m/n$. The relationship can be approximated by the power law $ \varphi_3 (m/n) = \left(63.4\right)^{m/n}$. (C) Fluvial energy expenditure vs. the channelization index. At large $\mathcal{C_I}$ values, the curves collapse on the same value determined by the ratio $m/n$.}
\end{figure}

To investigate the functional form of $\varphi_2(m,n)$ in Eq. \eqref{eq:phi_2}, we performed numerical simulations for non-unity values of $n$. The results reveal that the plateau value of $\hat{S}_b$ is uniquely related to the ratio $m/n$ (Fig. \ref{fig:m_n_ratio}A). The asymptotic value of $\hat{S}_b$ for $\ln{\mathcal{C_I}} \gg 1$ is a function of only one argument
\begin{equation}
\label{f_S_b_3}
    \hat{S_b} = \varphi_3\left(\frac{m}{n}\right),
\end{equation}
where the functional form of $\varphi_3$ is presented in Fig. \ref{fig:m_n_ratio}B.

This relationship between $\hat{S}_b$ and the ratio $m/n$ can be explained through the variational analysis of the LEM for the case of negligible diffusion \cite{hooshyar2020variational}. The steady-state ridge/valley topography has the optimal fluvial energy expenditure $\int_\Omega \hat{a}^{1-{m/n}} d\Omega$ as $\ln{\mathcal{C_I}} \gg 1$, where the integral is over the domain $\Omega$. Since the emerging spatial arrangement of ridges and valleys is only a function of $m/n$ in the fluvial erosion regime rather than the absolute values of these exponents (Fig. \ref{fig:m_n_ratio}C), the average sediment partitioning is controlled by the ratio $m/n$, following Eq. \eqref{f_S_b_3}.

The optimality principle for LEM, originally obtained in \cite{hooshyar2020variational}, is re-derived here more concisely using Euler-Lagrange equations with an explicit focus on $\ln{\mathcal{C_I}} \gg 1$. A constrained Lagrangian function in terms of $\hat{a}$ and its derivatives with imposed continuity (Eq. \eqref{eq:a_nd}) is written as
\begin{equation}
\mathcal{L}[\hat{a}, \hat{a}_{\hat{x}}, \hat{a}_{\hat{y}}] = \int_\Omega \left[ \hat{a}^{1-m/n} + \lambda \left( \hat{\nabla} \cdot \left(\hat{a}\frac{\hat{\nabla} \hat{z}}{| \hat{\nabla} \hat{z}|}\right) + 1 \right) \right] d \Omega,
\end{equation}
where $\lambda$ is a Lagrange multiplier field, and $\hat{a}_{\hat{x}}$ and $\hat{a}_{\hat{y}}$ are the partial derivatives of $\hat{a}$ with respect to $\hat{x}$ and $\hat{y}$, respectively.

Finding critical functions for which the first variation vanishes ($\delta \mathcal{L} = 0$) gives the Euler-Lagrange equations
\begin{eqnarray}
&& \frac{\partial \mathcal{L}}{\partial \hat{a}} - \frac{\partial}{\partial \hat{x}} \left( \frac{\partial \mathcal{L}}{\partial \hat{a}_{\hat{x}}} \right) - \frac{\partial}{\partial \hat{y}} \left( \frac{\partial \mathcal{L}}{\partial \hat{a}_{\hat{y}}} \right) = 0, \\
&& \left(1-\frac{m}{n}\right) \hat{a}^{-m/n} - \hat{\nabla} \lambda \cdot \frac{\hat{\nabla} \hat{z}}{\left| \hat{\nabla} \hat{z} \right|} = 0.\label{eq:J_first}
\end{eqnarray}
Substituting $\lambda = \left(1 - \frac{m}{n}\right) \hat{z}$ in Eq. \eqref{eq:J_first} yields the steady-state solution of Eq. \eqref{eq:z_nd} for negligible diffusion. This verifies that the steady landscapes eroded by surface dynamics exist in a state of stationary $\int_\Omega \hat{a}^{1-m/n} d\Omega$ as $\ln{\mathcal{C_I}} \gg 1$. Neglecting the global contribution of diffusion in the variational analysis allows a clear derivation of the optimality principle, which is similar to the insights obtained by neglecting viscous dissipation and employing the Euler equations instead of Navier-Stokes equations, to obtain the underlying variational principle \cite{mobbs1982variational}.

\begin{figure}[!hbt]
\centering
\includegraphics[width=\linewidth]{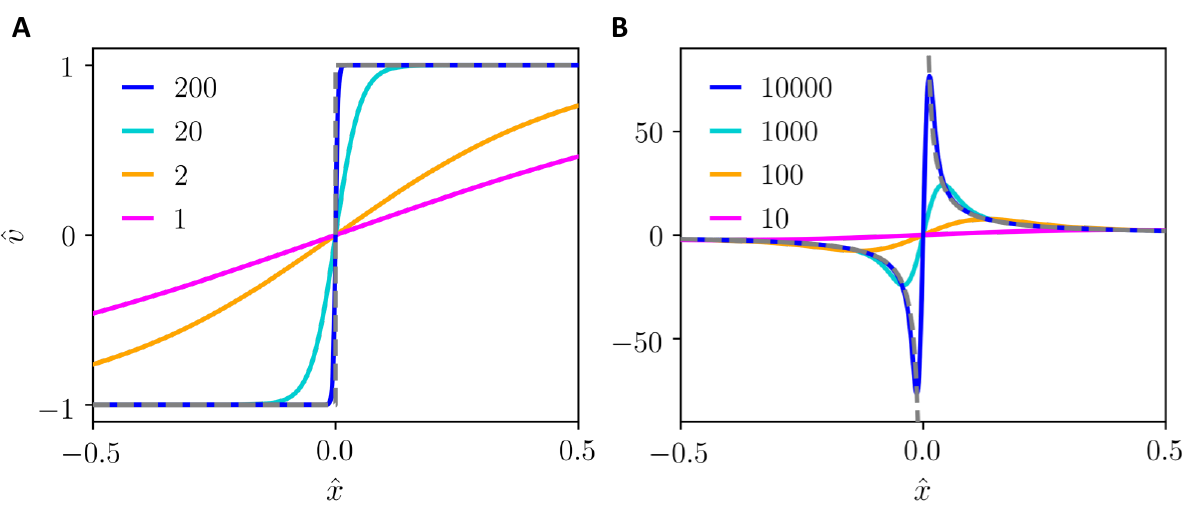}
\caption{\label{fig:v_SI}Vanishing diffusion limit of analytical 1D landscapes. The plot of $\hat{v} =-d\hat{z}/d\hat{x}$ for the exponents (A) $m=0,n=2$ and (B) $m=n=1$ at different channelization index values. The dashed curve represents the solution for vanishing soil diffusion, resulting in jump discontinuity for $m=0,n=2$ and infinite discontinuity for $m=n=1$.}
\end{figure}

\end{document}